\documentclass[prl,nofootinbib,reprint,longbibliography]{revtex4-2}

\usepackage{lipsum}
\usepackage{amsmath,amssymb,amsfonts,bbold}
\usepackage{braket}
\usepackage{natbib}
\usepackage{graphicx,subfigure}
\usepackage{hyperref}
\usepackage{mathtools}
\usepackage{comment}
\usepackage[normalem]{ulem}
\newcommand{\be}{\begin{equation}}
\newcommand{\ee}{\end{equation}}
\newcommand{\bse}{\begin{subequations}}
\newcommand{\ese}{\end{subequations}}
\newcommand{\ba}{\begin{eqnarray}}
\newcommand{\ea}{\end{eqnarray}}
\newcommand{\bea}{\begin{eqnarray}}
\newcommand{\eea}{\end{eqnarray}}

\newcommand{\lb}{\left (}
\newcommand{\rb}{\right )}
\usepackage{color}
\usepackage[normalem]{ulem}  

\begin{document}

% Use the \preprint command to place your local institutional report
% number in the upper righthand corner of the title page in preprint mode.
% Multiple \preprint commands are allowed.
% Use the 'preprintnumbers' class option to override journal defaults
% to display numbers if necessary
%\preprint{}

%Title of paper
\title{The degrees of freedom of multiway junctions in three dimensional gravity}

\author{Avik Chakraborty}
\email{avik.phys88@gmail.com}
\affiliation{Departamento de Ciencias F\'isicas, Facultad de Ciencias Exactas, Universidad Andres Bello,
Sazi\'e 2212, Piso 7, Santiago, Chile}
\author{Tanay Kibe}
\email{tanay.kibe@wits.ac.za }
\affiliation{National Institute for Theoretical and Computational Sciences,
School of Physics and Mandelstam Institute for Theoretical Physics,
University of the Witwatersrand, Wits, 2050, South Africa}
\author{Mart\'in Molina}
\email{martinmolinaramos95@gmail.com}
\affiliation{Instituto de F\'{\i}sica, Pontificia Universidad Cat\'{o}lica de Valpara\'{\i}so,
Avenida Universidad 330, Valpara\'{\i}so, Chile}
\affiliation{Departamento de F\'{\i}sica, Universidad T\'{e}cnica Federico Santa Mar\'{\i}a,
Casilla 110-V, Valpara\'{\i}so, Chile,}
\author{Ayan Mukhopadhyay}
\email{ayan.mukhopadhyay@pucv.cl}
\affiliation{Instituto de F\'{\i}sica, Pontificia Universidad Cat\'{o}lica de Valpara\'{\i}so,
Avenida Universidad 330, Valpara\'{\i}so, Chile}
\author{Giuseppe Policastro}
\email{giuseppe.policastro@phys.ens.fr}
\affiliation{Laboratoire de Physique de l'\'{E}cole Normale Supérieure, ENS, Universit\'{e} PSL, CNRS, Sorbonne Universit\'{e}, Universit\'{e} de Paris, F-75005 Paris, France}

\date{\today}

\begin{abstract}
We demonstrate that $n$-way junctions in three dimensional gravity correspond to coupled $n-1$ strings each satisfying the Nambu-Goto equation in the smoothened background, and with sources consisting of Monge-Amp\`{e}re like terms which couple the strings. For $n\geq 3$, these $n-1$ degrees of freedom survive the tensionless limit implying that matter-like behavior can arise out of \textit{pure} gravity. We interpret these stringy degrees of freedom of gravitational junctions holographically in terms of wavepackets which
collectively undergo perfect reflection at the multi-interface in the dual conformal field theory.
\end{abstract}

\maketitle
\textit{Introduction:-} Geometrizing matter provides a route towards understanding its fundamental origin. String theory concretely exemplifies this by demonstrating that both matter and gravity can emerge from the vibrations of a quantized fundamental string \cite{Green:2012oqa,Green:1987mn,Polchinski:1998rq,Polchinski:1998rr}. 
Alternatively, one can ask if matter-like degrees of freedom can emerge from gravity itself. This possibility can be probed by studying gravitational multi-way junctions that glue two \cite{Karch:2000ct,Karch:2001cw,DeWolfe:2001pq,Bachas:2001vj} or more \cite{Shen:2024dun,Shen:2024itl} spacetimes.
{In this letter, we consider} general multi-way gravitational junctions, both with and without tension, and {explore} if matter-like degrees of freedom can emerge even in the tensionless limit from \textit{pure} gravity. We generalize the earlier result showing that the non-linear Nambu-Goto equation for a fundamental string, with tension-induced corrections, emerges just from the junction conditions of a two-way junction gluing two three-dimensional Einstein manifolds \cite{Banerjee:2024sqq}. We demonstrate that junctions gluing three or more such spacetimes have non-trivial degrees of freedom corresponding to coupled Nambu-Goto equations with non-trivial solutions even in the tensionless limit. We also discuss extensions to four-dimensional settings.

The holographic correspondence \cite{Maldacena:1997re,Gubser:1998bc,Witten:1998qj}, which relates a strongly coupled quantum field theory to a classical gravitational theory in one higher dimension, implies that the gravitational junction gluing multiple three dimensional locally anti-de Sitter (AdS) spacetimes can be translated to an interface between quantum wires described by strongly coupled two-dimensional field theories \cite{Karch:2000ct,Karch:2001cw,DeWolfe:2001pq,Bachas:2001vj,Shen:2024dun,Shen:2024itl}. The presence of degrees of freedom of the gravitational junction imply that the interface can be viewed as a tunable quantum processor \cite{Chakraborty:2025dmc}. 
Therefore, the classical gravitational junction is of fundamental importance for a deeper understanding of how spacetime emerges from the quantum information structure of strongly interacting quantum field theories \cite{harlow2018tasi,Jahn:2021uqr,Chen:2021lnq,Kibe:2021gtw}. Furthermore, it opens new directions in the understanding of quantum information processing in many-body systems. Motivated by this, we focus on gravitational junctions gluing three-dimensional locally AdS spacetimes.

\begin{figure}
    \centering
    \includegraphics[width=\linewidth]{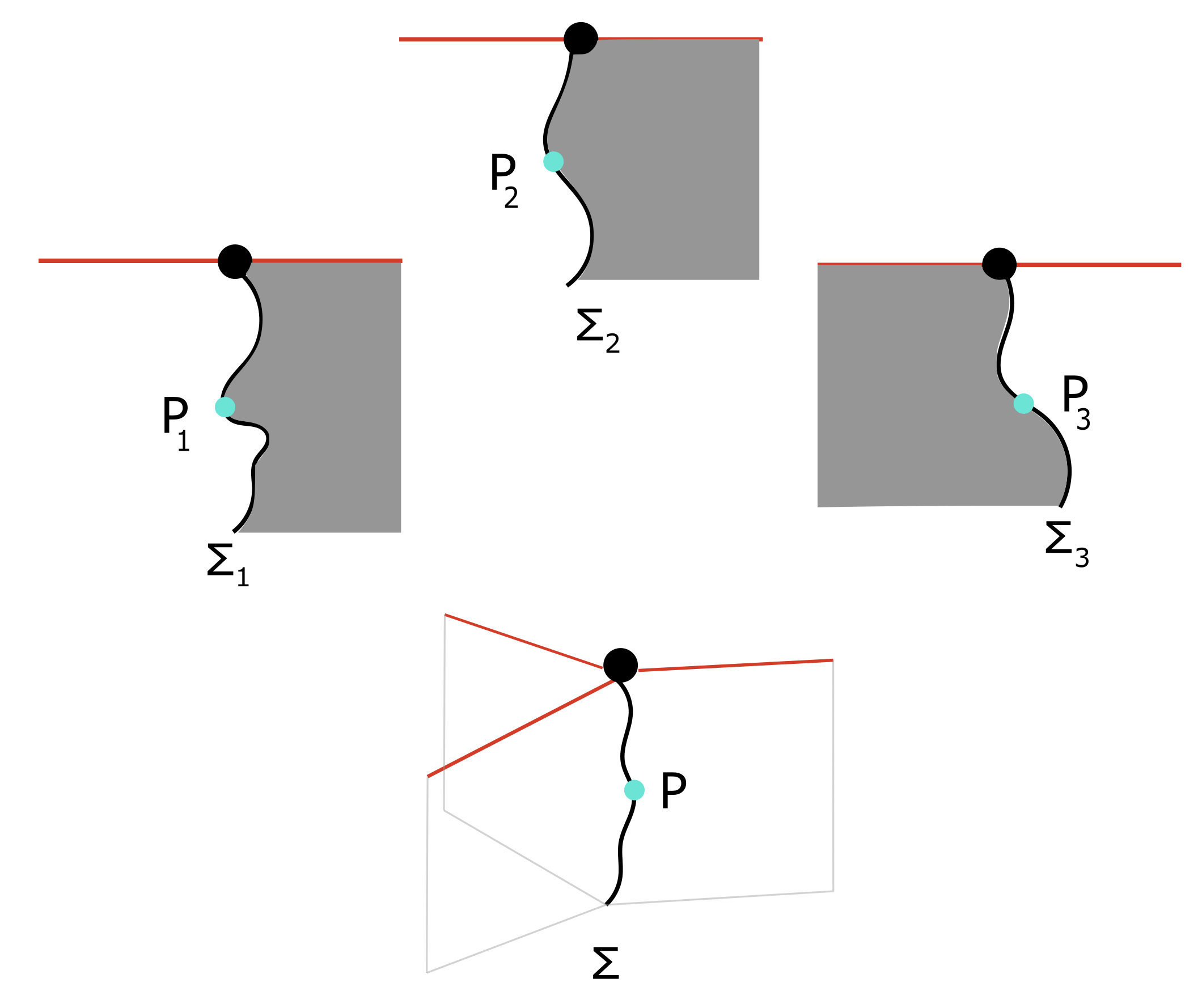}
    \caption{A three-way junction (below) gluing asymptotically AdS spaces (on top). Red line is the asymptotic boundary. Black dot is the interface in the dual CFT. The gray regions on left/right of $\Sigma_i$ ($i=1,2,3$) are excised. The  points $P_i$ on $\Sigma_i$ are identified with $P$ on $\Sigma$, the gravitational junction in $\widetilde{\mathcal{M}}$.}
    \label{fig:multiway}
\end{figure}

\textit{Gravitational multiway junctions:-} A  general three-dimensional Einstein spacetime $\mathcal{M}$ is locally flat, locally AdS$_3$ or locally dS$_3$. Consider $n$ identical copies, $\mathcal{M}_i$ of $\mathcal{M}$, each of which is divided into two parts, $\mathcal{M}_{iL}$ and $\mathcal{M}_{iR}$ along \textit{distinct} boundaries $\Sigma_i$. As each copy $\mathcal{M}_i$ inherits the coordinate charts of $\mathcal{M}$, we can clearly distinguish between the left and right halves $\mathcal{M}_{iL}$ and $\mathcal{M}_{iR}$ in each $\mathcal{M}_i$. Discarding $\mathcal{M}_{iL}$ or $\mathcal{M}_{iR}$ for each copy $i$, we glue the $n$ fragments $\mathcal{M}_{i\alpha_i}$ (with $\alpha_i = L,R$) to form a spacetime $\widetilde{\mathcal{M}}$ which satisfies Einstein's equations, including the necessary junction conditions \cite{Israel:1966rt}. The junction in $\widetilde{\mathcal{M}}$ is the co-dimension one hypersurface $\Sigma$, each point $P$ of which is formed by the identification of corresponding points $P_i$ in $\Sigma_i$. Therefore, each $\Sigma_i$ should be regarded as the image of $\Sigma$ in the corresponding $\mathcal{M}_i$. See Fig. \ref{fig:multiway}. The \textit{embeddings} of $\Sigma_i$ in $\mathcal{M}_i$ and the \textit{identifications} of the points $P_i$ of $\Sigma_i$ should be determined so that the junction conditions are satisfied at $\Sigma$. Our purpose is to investigate the nature of general solutions.
Before discussing the junction conditions explicitly, it is useful to set-up our notations and identify the variables that need to be determined. Let the coordinates of $\mathcal{M}$ be $t$, $z$ and $x$ (with $t$, the time coordinate, and $z$ and $x$ the spatial coordinates), and so the fragments $\mathcal{M}_{i\alpha_i}$ have coordinates $t_i$, $z_i$ and $x_i$. 
The boundaries $\Sigma_i$ of $\mathcal{M}_{i\alpha_i}$ that are glued at the junction $\Sigma$ assume the form
\begin{equation}\label{Eq:embed}
  \Sigma_i:\,\,  x_i = f_{i}(t_i, z_i), \,\, i = 1, 2,\cdots,n.
\end{equation}
We identify the points $P_i$ in $\Sigma_i$ with coordinates $(\tau_i,\sigma_i)$ with each point $P$ in $\Sigma$ with coordinates $(\tau,\sigma)$. We fix the (worldsheet) gauge freedom corresponding to the choice of coordinates $(\tau,\sigma)$ by
\begin{align}\label{Eq:WSgauge}
    \tau(P) = \frac{1}{n}\sum_{i=1}^n t_i(P_i), \quad \sigma(P) = \frac{1}{n}\sum_{i=1}^n z_i(P_i),
\end{align}
with $t_i(P_i)$ and $z_i(P_i)$ being the time and longitudinal coordinates of the points $P_i$ in $\Sigma_i$ that are identified with $P$. We note that above and in \eqref{Eq:embed}, $t_i$, $z_i$ and thus $x_i$ are functions of $\sigma$ and $\tau$, the worldsheet coordinates. After gauge-fixing \eqref{Eq:WSgauge} (up to residual worldsheet gauge symmetries to be discussed later), we are left with $2(n-1)$ independent variables $t_i - t_j$ and $z_i - z_j$ ($i\neq j$) which are the relative shifts of space and time as we move from $\mathcal{M}_{i\alpha_i}$ to $\mathcal{M}_{j\beta_j}$ across the junction $\Sigma$. Along with the $n$ embedding functions $f_i$ of $\Sigma_i$, we have in total $3n-2$ variables which need to be determined as functions of $\tau$ and $\sigma$.

The full gravitational action which gives the junction conditions is
\begin{align}\label{Eq:bulk-action}
   & S=\frac{1}{16\pi G_N}\int_{\mathcal{\widetilde{\mathcal{M}}}}d^3x \sqrt{-g}(R - 2\Lambda)+T_0\int_{\Sigma}{\rm d}\tau{\rm d}\sigma\,\sqrt{-\gamma}\nonumber\\&\quad+{\rm GHY \ terms}.
\end{align}
Note that the bulk metric $g$ is the \textit{only} degree of freedom in this action, and the second line consists of Gibbons-Hawking-York (GHY) terms corresponding to the boundaries of $\mathcal{M}_{i\alpha_i}$. The action \eqref{Eq:bulk-action} assumes that the induced metrics $\gamma_i$ on $\Sigma_i$ are identical at each point $P(\tau,\sigma)$ on $\Sigma$, thus defining the worldsheet metric via
\begin{align}\label{Eq:hcont}
&\gamma_{\mu\nu}(\tau, \sigma) := \gamma_{1,\mu\nu}(\tau,\sigma)=\cdots = \gamma_{n,\mu\nu}(\tau,\sigma).
\end{align}
Varying the action with respect to $g$ away from the junction, we obtain that each fragment $\mathcal{M}_{i\alpha_i}$ is an Einstein manifold. The variation with respect to $g$ at the junction gives 
%{\gp we should specify the convention for the normals}
\begin{equation}\label{Eq:Kdisc}
    \sum_{i=1}^n (-1)^{s(\alpha_i)}\left(K_{i,\mu\nu} - K_i\,\gamma_{i,\mu\nu}\right) = 8\pi G_N T_0 \gamma_{\mu\nu},
\end{equation}
with $s(\alpha_i) = 0$ if $\alpha_i = L$ and $s(\alpha_i) = 1$ if $\alpha_i = R$ (in each fragement $x<0$ and $x>0$ denotes the left and right halves, respectively). Above, $K_{i,\mu\nu}$ is the extrinsic curvature of $\Sigma_i$ in $\mathcal{M}_{i\alpha_i}$ and $K_i = \gamma^{\mu\nu}K_{i,\mu\nu}$. The bulk diffeomorphism symmetry implies the conservation of the \textit{total} Brown-York tensor of the junction, which is the left hand side of \eqref{Eq:Kdisc}. Therefore, the above equation has only one independent component.

{ We have $3n-3$ independent equations from \eqref{Eq:hcont} and only one from \eqref{Eq:Kdisc}. Thus the total number of independent junction conditions is $3n -2$, matching the total number of variables \textit{exactly}. 
We will show that the general solutions are in one-to-one correspondence with appropriately coupled Nambu-Goto equations and for $n\geq 3$, we can also obtain dynamical variables in the tensionless limit implying that matter like degrees of freedom can emerge from pure gravity.

Generalization of our results to higher dimensions is not straightforward because the number of independent junction conditions for the (co-dimension one) junction exceeds the number of variables, precluding emergent degrees of freedom in this case. Nevertheless, matter like degrees of freedom in the form of coupled strings can emerge also from pure gravity in  higher dimensions via embedding of the multiway junction between three-dimensional spaces. See the End Matter for more details.

In order to solve the dynamics of the gravitational junctions in  $3D$, it is useful to appropriately parameterize the $3n-2$ variables. For simplicity, let us first assume that $\alpha_i =L$ for all $i$. We define $3n-3$ independent relative shifts of the time ($\tau_{d_i}$), the longitudinal coordinate ($\sigma_{d_i}$) and the transverse coordinate ($x_{d_i}$) across the junction as follows:
\begin{eqnarray}\label{Eq:xsxd}
\tau_{d_i} &=&\begin{cases}
        \frac{1}{n}(t_n-t_{i+1}) \,\, {\rm for}\,\, i = 1, \cdots, n-2\\
        \frac{1}{n}(t_n-t_1) \,\, {\rm for}\,\, i = n-1
    \end{cases},\nonumber\\
\sigma_{d_i} &=&\begin{cases}
        \frac{1}{n}(z_n-z_{i+1}) \,\, {\rm for}\,\, i = 1, \cdots, n-2\\
        \frac{1}{n}(z_n-z_1) \,\, {\rm for}\,\, i = n-1
    \end{cases},\nonumber\\
x_{d_i} &=&\begin{cases}
        \frac{1}{n}(x_n-x_{i+1}) \,\, {\rm for}\,\, i = 1, \cdots, n-2\\
        \frac{1}{n}(x_n-x_1) \,\, {\rm for}\,\, i = n-1
    \end{cases}.
\end{eqnarray}

The above and the averaged transverse coordinate
\begin{equation}
    x_s = \frac{1}{n}\sum_i x_{i}.
\end{equation} 
give the requisite $3n-2$ functions of $\tau$ and $\sigma$ that need to be determined.

If a subset of the  $n$ fragments, $\mathcal{M}_{i\alpha_i}$ are $\mathcal{M}_{iR}$ instead of $\mathcal{M}_{iL}$, we simply reverse the sign of the transverse coordinate $x_i$ in the parameterization \eqref{Eq:xsxd} for the values of $i$ in this subset.

For $n=2$, the general solutions of the junction conditions in three dimensional gravity \cite{Banerjee:2024sqq} are in one-to-one correspondence with the solutions of Nambu-Goto equation in $\mathcal{M}$ up to six rigid parameters which are related to worldsheet and spacetime isometries (to be described later).  If we glue $\mathcal{M}_{1L}$ and $\mathcal{M}_{2R}$, then for a generic solution of the junction conditions:
\begin{enumerate}
    \item the hypersurface $$\Sigma_{NG}: t=\tau, \,\, z=\sigma,\,\, x = x_s(\tau,\sigma)$$in $\mathcal{M}$ (whose embedding is the average of $\Sigma_1$ and $\Sigma_2$) corresponds to a solution of the non-linear Nambu-Goto equations for a worldsheet in $\mathcal{M}$ when the tension $T_0$ and the rigid parameters vanish, and
    \item $x_s$ is the \textit{only} degree of freedom implying that $x_d$, $\tau_d$ and $\sigma_d$ are completely determined as functions of $\tau$, $\sigma$, the tension and the rigid parameters for any given choice of the solution of the Nambu-Goto equation in $\mathcal{M}$ corresponding to $x_s$.
\end{enumerate}
If we glue $\mathcal{M}_{1R}$ and $\mathcal{M}_{2R}$ (or $\mathcal{M}_{1L}$ and $\mathcal{M}_{2L}$), then the roles of $x_s$ and $x_d$ are reversed. 
%{\textcolor{red}{In this case, $x_d$ corresponds to a solution of the non-linear Nambu-Goto equation in $\mathcal{M}$, and $x_s$ along with $\tau_d$ and $\sigma_d$ are determined by $x_d$.} {\gp I think the last sentence may be delted in case we need more space.}  
In what follows, we will generalize these results and the perturbative analysis in \cite{Banerjee:2024sqq} for $n\geq 3$. We will focus our discussion on the $n$-way junction in AdS$_3$ with a view to its holographic interpretation.

\textit{Perturbative Analysis:-} Let us first consider $n\geq 3$ fragments of locally AdS$_3$ spacetimes, each of which is a copy of the Ba\~nados-Teitelboim-Zanelli (BTZ) black brane \cite{Banados:1992wn,Banados:1992gq} ($\mathcal{M}$) 
%{\gp we should motivate why we consider BTZ and not pure AdS} endowed with the metric
\begin{equation}
    {\rm d}s^2 = \frac{\ell^2}{z^2}\left(-(1-M z^2){\rm d}t^2 + \frac{{\rm d}z^2}{1-M z^2}+ {\rm d}x^2\right)
\end{equation}
which is dual to the thermal state of the dual CFT (more on this later). For convenience, we set $\ell =1$. We define $\lambda = 8\pi G_N T_0$ and consider $\lambda = \mathcal{O}(\epsilon)$. We solve the general junction conditions perturbatively in $\lambda$ for the gluing of the $n$ fragments $\mathcal{M}_{iL}$. At the zeroth order, we choose the simplest (permutation-symmetric) solution
\begin{align}
    &x_i = x_0 +\mathcal{O}(\epsilon) \Rightarrow x_s = x_0 +\mathcal{O}(\epsilon),\,\, x_{d_i} =\mathcal{O}(\epsilon), \nonumber\\&{\rm and} \,\,\,\tau_{d_i} =\sigma_{d_i} =\mathcal{O}(\epsilon^2),
\end{align}
so that we find at least one static solution to all orders in $\lambda$ which has the interpretation of a \textit{conformal} defect joining $n$ copies of the dual CFT living on semi-infinite lines (wires). 

At the first order in $\epsilon$, the solutions for $\tau_{d_i}$ and $\sigma_{d_i}$ correspond to the three worldsheet Killing vectors whereas the solution for $x_s - x_0$ feature the three spacetime isometries of $\mathcal{M}$ which do not preserve the zeroth order hypersurface $x_s = x_0$. In what follows, we set these $3n$ rigid parameters to zero. We also look for solutions in which
\begin{equation}\label{Eq:DBC}
    \lim_{\sigma\rightarrow 0}x_i = x_0\Rightarrow \lim_{\sigma\rightarrow 0}x_s = x_0, \,\,\lim_{\sigma\rightarrow 0}x_{d_i} =0
\end{equation}
corresponding to the Dirichlet boundary condition at the boundary of AdS. Such solutions have the interpretation of a $n$ way interface in the dual CFT. 
%{\gp It is not completely general, even in a two-way junction there are other interfaces, for instance the ones considered by Karch with multiple branes ending on the same line at the boundary.}

We find that 
\begin{equation}\label{Eq:xssol}
    x_s = x_0+\frac{\lambda }{n}\sigma + \mathcal{O}(\epsilon^3).
\end{equation}

Remarkably, the $n-1$ variables $x_{d_i}$ correspond to coupled Nambu-Goto equations in $\mathcal{M}$.  Up to third order in $\epsilon$, these explicitly are
\begin{align}\label{Eq:CoupledNGEOMs}
    &\lambda \, \mathcal{N}_i = n\left((n-2) \mathcal{X}_i -\sum_{j\neq i}\mathcal{Z}_{ij} \right)+\mathcal{O}(\epsilon^3),
\end{align}
for $i=1,\cdots ,n-1$, with  primes and dots denoting $\partial_\sigma$ and $\partial_\tau$, respectively, and
\begin{align}\label{Eq:CNG}
   & \mathcal{N}_i=(1-M\sigma^2)\Big((1-M\sigma^2)\Big(-2x^{\prime}_{d_i}+\nonumber\\&\qquad\quad +\sigma x^{\prime \prime}_{d_i}(1-M\sigma^2)\Big) - \sigma \ddot{x}_{d_i} \Big)
\end{align}
corresponding to the linearized Nambu-Goto equation for the hypersurface
\begin{equation}
    \Sigma_{NG_i}: t=\tau, \,\, z=\sigma,\,\, x = x_{d_i}(\tau,\sigma)
\end{equation}
in $\mathcal{M}$, while
\begin{eqnarray}
    \mathcal{X}_i&=&x^{\prime 2}_{d_i} (1-M\sigma^2)^2\nonumber\\&&+ \sigma x^{\prime}_{d_i}(1-M\sigma^2) (-(1-M\sigma^2)^2 x^{\prime \prime}_{d_i} + \ddot{x}_{d_i} )\nonumber\\&&+ \sigma^2((M\sigma \dot{x}_{d_i}+(1-M\sigma^2)\dot{x}^{\prime }_{d_i})^2 
    %\nonumber\\&&\qquad\quad\quad
    \nonumber\\&&
    -(1-M\sigma^2)^2 x^{\prime \prime}_{d_i}\ddot{x}_{d_i}),
\end{eqnarray}
and
\begin{eqnarray}
    \mathcal{Z}_{ij}&=&2x^{\prime}_{d_i}x^{\prime}_{d_j} (1-M\sigma^2)^2\nonumber\\&&+ \sigma x^{\prime}_{d_i}(1-M\sigma^2) (-(1-M\sigma^2)^2 x^{\prime \prime}_{d_j} + \ddot{x}_{d_j} )\nonumber\\&&+ \sigma x^{\prime}_{d_j}(1-M\sigma^2) (-(1-M\sigma^2)^2 x^{\prime \prime}_{d_i} + \ddot{x}_{d_i} )\nonumber\\&&+ \sigma^2(2(M\sigma \dot{x}_{d_i}+(1-M\sigma^2)\dot{x}^{\prime }_{d_i})\nonumber\\&&\qquad \qquad\times(M\sigma \dot{x}_{d_j}+(1-M\sigma^2)\dot{x}^{\prime }_{d_j})
    %\nonumber\\&&\qquad\quad\quad
    \nonumber\\&&
    -(1-M\sigma^2)^2 (x^{\prime \prime}_{d_i}\ddot{x}_{d_j}+ x^{\prime \prime}_{d_j}\ddot{x}_{d_j})),
\end{eqnarray}
are Monge-Amp\`ere like terms coupling the Nambu-Goto equations.  Notice that the rhs of \eqref{Eq:CoupledNGEOMs} vanishes for $n=2$.

The physical solutions corresponding to a holographic interface of $n$ wires should satisfy the Dirichlet boundary conditions \eqref{Eq:DBC}. While $x_s$ is given by \eqref{Eq:xssol}, the solutions of $x_{d_i}$ with such Dirichlet boundary conditions are of different nature when $\lambda >0$ as opposed to $\lambda =0$ for $n\geq 3$, as turning on $\lambda$ is a singular perturbation of \eqref{Eq:CoupledNGEOMs}. 
(For $n=2$, the equation for $x_d$ trivially vanishes when $\lambda =0$.) For illustration, consider $n=3$. In this case, when $\lambda \neq 0$, the solutions of \eqref{Eq:CoupledNGEOMs} to $\mathcal{O}(\sigma^7,\epsilon^2)$ are
\begin{align}
    x_{d_1} &= \mathcal{A}_1 \sigma^3
+ \frac{3\mathcal{A}_1 \left(2 M \lambda + 9\mathcal{A}_1 - 18\mathcal{A}_2 \right) + \lambda \ddot{\mathcal{A}}_1}{10 \lambda} \sigma^5,\nonumber \\
x_{d_2} &= \mathcal{A}_2 \sigma^3
+ \frac{3\mathcal{A}_2 \left(2 M \lambda - 18\mathcal{A}_1 + 9\mathcal{A}_2 \right) + \lambda \ddot{\mathcal{A}}_2}{10 \lambda} \sigma^5 ,
\end{align}
where the normalizable modes $\mathcal{A}_1(\tau)$ and $\mathcal{A}_2(\tau)$, which are both $\mathcal{O}(\epsilon)$ (like $\lambda$), determine the full solutions; and when $\lambda =0$, the analogous solutions of \eqref{Eq:CoupledNGEOMs} are
\begin{widetext}
\begin{align}
    x_{d_1} &= \mathcal{B}_1 \sigma^2  +\frac{
\mathcal{B}_1 \left( M \left( \mathcal{B}_1^2 - \mathcal{B}_1 \mathcal{B}_2 + \mathcal{B}_2^2 \right) - \dot{\mathcal{B}}_2^2 \right)
+ 2 \mathcal{B}_2 \dot{\mathcal{B}}_1 \dot{\mathcal{B}}_2 
+ (\mathcal{B}_1 - \mathcal{B}_2) \dot{\mathcal{B}}_1^2
}{
4 \left( \mathcal{B}_1^2 - \mathcal{B}_1 \mathcal{B}_2 + \mathcal{B}_2^2 \right)
} \sigma^4  + \mathcal{O}(\sigma^6,\epsilon^2),\notag \\
x_{d_2} &= \mathcal{B}_2 \sigma^2+\frac{
\mathcal{B}_2 \left( M \left( \mathcal{B}_1^2 - \mathcal{B}_1 \mathcal{B}_2 + \mathcal{B}_2^2 \right) - \dot{\mathcal{B}}_1^2 \right)
+ 2 \mathcal{B}_1 \dot{\mathcal{B}}_1 \dot{\mathcal{B}}_2 
+ (-\mathcal{B}_1 + \mathcal{B}_2) \dot{\mathcal{B}}_2^2
}{
4 \left( \mathcal{B}_1^2 - \mathcal{B}_1 \mathcal{B}_2 + \mathcal{B}_2^2 \right)
} \sigma^4 + \mathcal{O}(\sigma^6,\epsilon^2),
\end{align}
\end{widetext}
where the normalizable modes $\mathcal{B}_1(\tau)$ and $\mathcal{B}_2(\tau)$, both of which are $\mathcal{O}(\epsilon)$, determine the full solutions. Generally, when the Dirichlet boundary conditions \eqref{Eq:DBC} are imposed, $x_{d_i}$ are even in $\sigma$ and vanish as $\sigma^2$ near the boundary for $\lambda =0$, while they are odd in $\sigma$ and vanish as $\sigma^3$ near the boundary when $\lambda\neq 0$. For the holographic interpretation of the multiway junction, the most relevant aspect of $\tau_{d_i}$ is that
\begin{equation}
\lim_{\sigma\rightarrow0}\tau_{d_i}(\sigma,\tau) = \mathbb{t}_{d_i}(\tau), \quad i= 1,\cdots,n-1
\end{equation}
are non-vanishing indicating that the relative time differences (reparameterizations) are non-trivial at the dual interface. Furthermore, $\mathbb{t}_{d_i}(\tau)$ are determined by the normalizable modes of $x_{d_i}$. When $\lambda \neq0$,
\begin{equation}\label{Eq:TR1}
    {\dddot{\mathbb{t}}_{d_i}} - M \dot{\mathbb{t}}_{d_i} = -\frac{3}{n}\lambda\mathcal{A}_i + \mathcal{O}(\epsilon^3),\,\, i = 1,\cdots,n-1,
\end{equation}
for $n\geq 2$, and when $\lambda =0$,
\begin{equation}\label{Eq:TR2}
    {\dddot{\mathbb{t}}_{d_i}} - M \dot{\mathbb{t}}_{d_i} = 2\mathcal{B}_i\left((n-2)\mathcal{B}_i - 2\sum_{j\neq i}\mathcal{B}_j\right)+ \mathcal{O}(\epsilon^3)
\end{equation}
for $n\geq 3$. In order to eliminate rigid transformations related to worldsheet isometries mentioned before, we require that $\mathbb{t}_{d_i}$ decay faster than $e^{-\sqrt{M}\tau}$ at large $\tau$. (It can be checked that $\mathcal{A}_i$ decay as $e^{-m\sqrt{M}\tau}$ with $m= 2,3,\cdots$ at large $\tau$ for $x_{d_i}$ to be ingoing at the worldsheet horizon. \footnote{ We intend to study the behavior of the $\mathcal{B}_i(\tau)$ modes explicitly in the future. Note that the equations \eqref{Eq:CoupledNGEOMs} cannot be linearized when $\lambda=0$.}) As in the case of $n=2$, $\tau_{d_i}$ and $\sigma_{d_i}$ are also then completely determined by the normalizable modes of $x_{d_i}$.

Crucially, $\mathbb{t}_{d_i}$, the relative time reparameterizations at the dual interface completely encode the normalizable modes of $x_{d_i}$, and therefore the solution of the coupled Nambu-Goto equations for $x_{d_i}$ which constitute the $n-1$ degrees of freedom of the gravitational multiway junction. When $n\geq 3$, there are non-trivial degrees of freedom even when $\lambda =0$ which can be decoded from the relative time reparameterizations $\mathbb{t}_{d_i}$ at the dual interface.

\textit{Holographic interpretation:-} We can proceed with the holographic interpretation of gravitational junctions in terms of interfaces of dual large $c$ conformal field theories {(CFT)} by applying the holographic dictionary \cite{Henningson:1998gx,Balasubramanian:1999re}. The simplest solution in which $x_{d_i}$, $t_{d_i}$ and $\sigma_{d_i}$ vanish corresponds to a static conformal interface in the dual CFT where $n$ thermal wires with equal temperature are glued. The tension $\lambda$ characterizes the defect operator at the interface  e.g. via the boundary entropy \cite{PhysRevLett.67.161,Azeyanagi:2007qj,Afxonidis:2024gne}. 

For the holographic interpretation of the general solutions of the gravitational junction in terms of physical operations, we note that the relative time reparameterizations $\mathbb{t}_{d_i}$ (encoding the normalizable modes of $x_{d_i}$) can be undone by conformal transformations acting on $n-1$ of the $n$ wires glued at the dual interface. Generalizing the case of $n=2$ analyzed in \cite{Chakraborty:2025dmc}, we can proceed by first interchanging $\tau $ with $t_n(\sigma =0, \tau)$, the time coordinate of the $n^{\rm th}$ wire at the boundary, and then $\mathbb{t}_{d_i}$ with
\begin{equation}\label{Eq:timereparam}
    t_i = \mathbb{h}_i(t_n), \,\, i = 1,\cdots,n-1
\end{equation}
at $\sigma =0$, relating the time of the $i$th wire to that of the $n$th wire at the interface, which can be located at $x_i= x_0 =0$ without loss of generality (at the boundary). Let $x^\pm_i = t_i\pm x_i$ be the lightcone coordinates of the wires. The $n-1$ conformal transformations with
\begin{equation}\label{Eq:CTs}
    \tilde{x}^\pm_i = \mathbb{h}_i^{-1}(x^\pm_i), \,\, i = 1,\cdots,n-1
\end{equation}
give new coordinates $\tilde{x}_i= (1/2)(\tilde{x}_i^+ -\tilde{x}_i^-)$  and $\tilde{t}_i= (1/2)(\tilde{x}_i^+ +\tilde{x}_i^-)$. We note from \eqref{Eq:CTs}, that the position of the interface is preserved at $\tilde{x}_i =0$ while $\tilde{t}_i = t_n$ at the interface for $i=1,\cdots,n-1$. Therefore, we obtain continuous coordinates and a continuous metric across the interface as a result of the conformal transformations \eqref{Eq:CTs}. However, the energy-momentum tensor becomes discontinuous at the interface with the following non-vanishing components for $i=1,\cdots,n-1$
{\begin{equation}\label{Eq:Tmni}
   \widetilde{T}_{\pm\pm}^i(\tilde{x}_i^\pm) =\frac{\pi c}{12} \mathbb{h}_i'(\tilde{x}_i^\pm)^2 T^2 -\frac{c}{24\pi}{\rm Sch}(\mathbb{h}_i(\tilde{x}_i^\pm),\tilde{x}_i^\pm).
\end{equation}}
 In the bulk, the conformal transformations can be lifted to (improper) bulk diffeomorphisms \cite{deHaro:2000vlm} on each of the $n-1$ fragments $\mathcal{M}_{i\alpha_i}$ for $i=1,\cdots,n-1$, and the holographic dictionary reproduces \eqref{Eq:Tmni} with $c =\frac{3\ell}{2G_N}$ and the temperature $T=\frac{\sqrt{M}}{2\pi}$ \cite{Balasubramanian:1999re,Henningson:1998gx,deHaro:2000vlm}. As the induced metric of a hypersurface and its extrinsic curvature transform as scalars under bulk diffeomorphism, we obtain a gauge equivalent solution of the gravitational junction. It is to be noted that the conformal transformations \eqref{Eq:CTs} do not involve any rigid SL(2,R) transformation which preserves the thermal state as we discarded the corresponding rigid parameters in the gravitational solution (involving homogeneous solutions of \eqref{Eq:TR1} or \eqref{Eq:TR2}). For a statement of Ward identities of the dual interface, see the End Matter. 

The solution of the coupled Nambu-Goto equations given by the $n-1$ normalizable modes of $x_{d_i}$ determines the functions $\mathbb{h}_i$. Thus, the $n-1$ degrees of freedom of the gravitational junction translate to state-dependent wave packets on $n-1$ wires converging to the interface collectively and reflected \textit{without} distortion at the interface individually, as implied by \eqref{Eq:Tmni}, when all the wires are at the same background temperature. The state-dependence is a consequence of the fact that the functions $\mathbb{h}_i$ encode the solution of the coupled Nambu-Goto equations in the BTZ black brane background which is dual to the thermal state. 

\textit{Conclusions:-} {We have shown that gravitational multiway junctions gluing $n$ copies of BTZ black branes correspond to $n-1$ coupled Nambu-Goto equations, implying emergent matter like degrees of freedom from gravity, which survive in the limit in which the tension of the junction is taken to zero for $n\geq 3$. Given that we obtain a non-smooth spacetime manifold with emergent matter-like degrees of freedom even in the tensionless limit, our solutions are entirely novel in the context of pure gravity in three dimensions. As discussed in the End Matter, it is possible to generalize our results to higher dimensional pure gravity.

We have also shown that a $n$-way gravitational junction gluing $n$ copies of BTZ black branes is holographically dual to an interface of $n$ wires at the same temperature, each described by a large $c$ strongly interacting two-dimensional CFT. The modes described by coupled Nambu-Goto equations correspond to possible combinations of wavepackets on the thermal wires, which can scatter off without distortion at the dual interface from past null infinity to future null infinity at each wire individually.} 

Extending our constructions to junctions gluing multiple \textit{dissimilar} locally AdS$_3$ spacetimes is a problem of fundamental importance. Deformations of the wavepackets which scatter off at the dual interface without distortion correspond to the deformations of the corresponding background geometries from BTZ black branes to general (locally AdS$_3$) Ba\~nados geometries \cite{Banados:1998gg}, and should lead to $n\rightarrow n$ scattering with tunable energy transmissions/reflections/re-distributions. As in the case of the two-way junction \cite{Bachas:2020yxv,Chakraborty:2025dmc,Banerjee:2025zuw}, the stringy degrees of freedom can be reconstructed as  universal quantum maps. Crucially, our results together with those discussed in \cite{Chakraborty:2025dmc,Banerjee:2025zuw} indicate that the multiway interfaces of CFTs \cite{Karch:2000ct,Karch:2001cw,DeWolfe:2001pq,Bachas:2001vj,Lap:2024vwm} can be generalized by involving relative automorphisms of the Virasoro algebra, and can be realized in spin and quantum Hall systems \cite{fendley1995exact,Oshikawa:1996dj,fal1999topological,Oshikawa:2005fh,Gromov:2016umy}.

{Furthermore, holographic bulk spacetime reconstruction is best understood as the recovery map in a quantum error correcting code encoding bulk sub-regions to boundary subregions \cite{harlow2018tasi,Jahn:2021uqr,Chen:2021lnq,Kibe:2021gtw}. The bulk reconstruction of the emergent matter-like degrees of freedom of gravitational junctions from sub-regions of the dual dynamic multi-way interface poses a fundamental challenge within the holographic framework. Primarily, we should understand how these modes affect entanglement at the dual interface following the methodology of \cite{Azeyanagi:2007qj,Afxonidis:2024gne,Afxonidis:2025jph,Kibe:2021qjy,Banerjee:2022dgv,Kibe:2024icu}.}
%{\gp mention something about energy transmission?}

\begin{acknowledgments}
\textit{Acknowledgments:-} We especially thank Cheng Peng as discussions between him and AM motivated the present research. AC, AM and MM acknowledge support from FONDECYT postdoctoral grant no. 3230222, FONDECYT regular grant no. 1240955 and ``Doctorado Nacional'' grant no. 21250596 of La Agencia Nacional de Investigaci\'{o}n y Desarrollo (ANID), Chile, respectively. AC also appreciates the warm hospitality extended by AM and Instituto de F\'{\i}sica, Pontificia Universidad Cat\'{o}lica de Valpara\'{\i}so, Chile where majority of the work was carried out. AM gratefully acknowledges the hospitality of LPENS, where a substantial part of this work was carried out during his tenure as a CNRS invited professor.
\end{acknowledgments}

%\newpage
\section*{End Matter}
\subsection{Generalizations for $D>3$}
Let us first see how the count of the number of variables and number of independent junction conditions generalize for multiway co-dimension one gravitational junctions in dimensions $D>3$.

If we join $n$ fragments of $D$-dimensional spacetimes with $D\geq 3$, the total number of variables in an arbitrary gluing can be counted as follows. Firstly, we have $n$ embedding functions giving the transverse coordinate of the co-dimension one boundaries $\Sigma_i$ (like \eqref{Eq:embed}) and secondly we have $(D-1)(n-1)$ independent relative spacetime shifts in time and the $D-2$ longitudinal coordinates across the junction (we can identify the average time and longitudinal coordinates of the images $\Sigma_i$ of $\Sigma$ with the $D-1$ worldsheet coordinates like in \eqref{Eq:WSgauge}).  In total there are thus $$(D-1)(n-1) +n=Dn-D +1$$variables. The junction conditions involve $D(D-1)(n-1)/2$ conditions involving continuity of the induced metric and $D(D-1)/2 - (D-1) = (D-1)(D-2)/2$ independent conditions involving the total Brown-York tensor after accounting for the $(D-1)$ constraints implying its conservation. (The constraints originate from the symmetry of the action \eqref{Eq:bulk-action} with respect to bulk diffeomorphisms.) Therefore, the number of independent junction conditions total $$\frac{D(D-1)(n-1)}{2}+\frac{(D-1)(D-2)}{2}=\frac{D-1}{2}\left(Dn-2\right).$$We note that $$\frac{D-1}{2}\left(Dn-2\right)\geq Dn-D+1$$for $D\geq 3$ with the inequality saturated iff $(D-1)/2 =1$, i.e. $D=3$. Therefore, we do not expect degrees of freedom to emerge at co-dimension one gravitational junctions in $D>3$.

However, we can consider the embedding of the multi-way junctions gluing slices of three dimensional manifolds in $D = 3+k$ and our results suggest that matter-like degrees of freedom in the form of vibrating strings can emerge at these co-dimension $k+1$ junctions. This is more easily implementable for flat spacetimes where our results for the multi-way junctions gluing three dimensional spacetimes remain valid. Consider a tessellation of $\mathbb{R}^{3,1}$ by four dimensional hypercubes each bounded by six $2+1$ dimensional faces as shown in Fig. \ref{fig:higherdim}. The $1+1$ dimensional edges of the hypercubes are junctions where four three dimensional spacetimes are glued. This embeds the four way junction gluing four $2+1$ dimensional spacetimes studied in this letter in $\mathbb{R}^{3,1}$, providing a concrete setup where we can investigate whether string-like degrees of freedom can emerge even in \textit{pure} gravity in four dimensions. Similarly, we can proceed with polygonal tessellations of AdS$_4$. We postpone the study of such gravitational junctions in $D>3$ and the analysis of their holographic interpretations for the future.  
\begin{figure}
    \centering
    \includegraphics[width=0.5\linewidth]{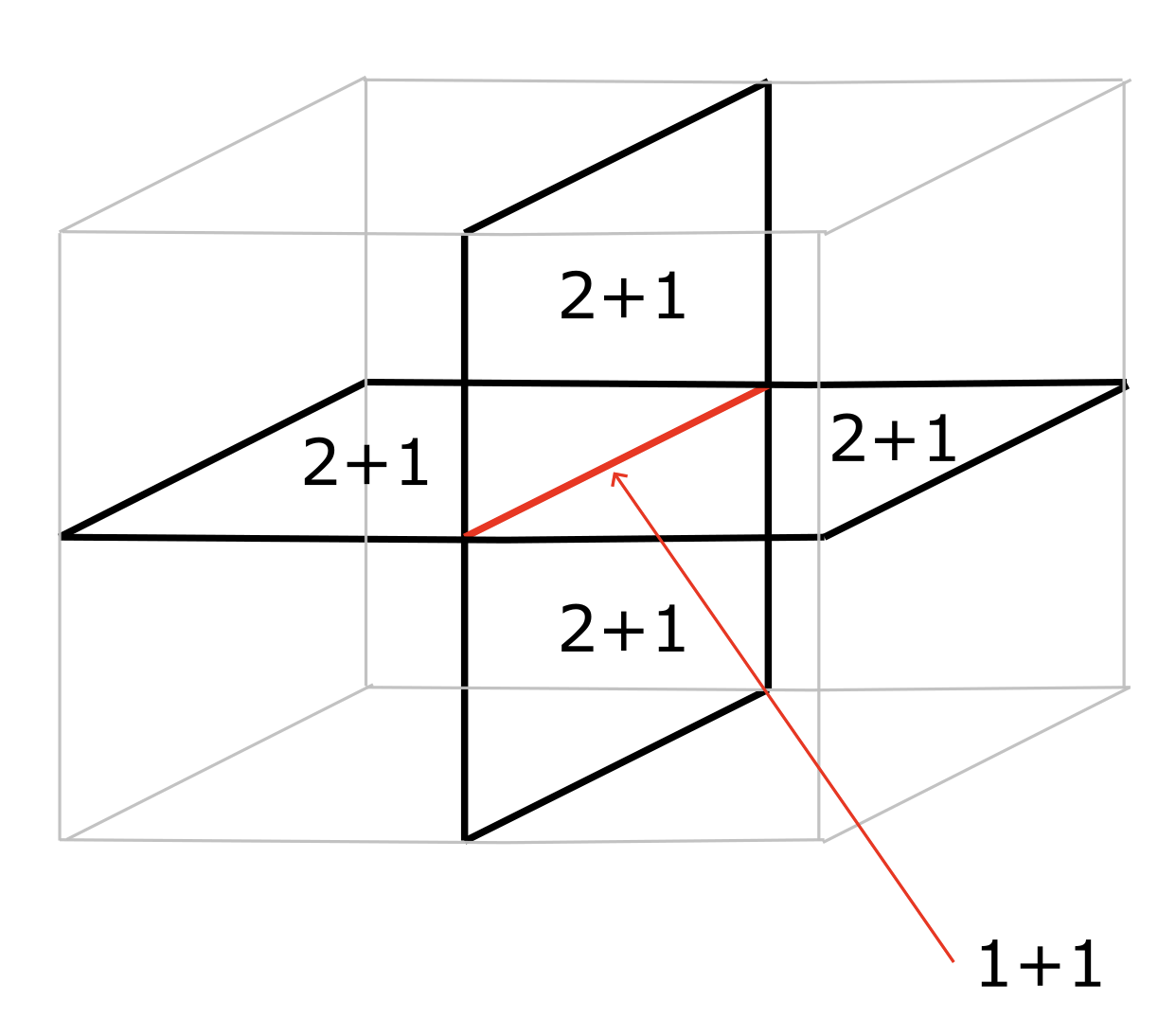}
    \caption{A sketch of the junction in $3+1$ dimensions. The four planes with solid black boundaries are $2+1$ dimensional and bound $3+1$ dimensional hypercubes (gray boundaries). Red line is a $1+1$ dimensional four-way junction.}
    \label{fig:higherdim}
\end{figure}
\subsection{Ward identities at the dual multi-interfaces}
Consider the $n-$way junction studied in this letter. In the dual $1+1$ dimensional CFT, it corresponds to $n$ wires joined at a single point, say $x_i=0$, with relative time reparameterizations between the wires. Let us fix the time coordinate of the $n^{\rm th}$ wire $t_n$ to be the global time. As shown in \eqref{Eq:timereparam} in the main text, the relative  time-reparametrization between the  $i^{\rm th}$ and $n^{\rm th}$ wire can be expressed as
\begin{equation}
    t_i = \mathbb{h}_i(t_n),\quad i=1,\cdots,n-1,
\end{equation}
and we can undo these time reparameterizations by using different conformal transformations on each of the $i\neq n$ wires. These transformations are
\begin{equation}
    \tilde{x}_i^\pm = \mathbb{h}^{-1}_i(x^\pm_i), \quad i=1,\cdots, n-1
\end{equation}
where $x_i^\pm= t_i\pm x_i$ are the lightcone coordinates for each of the wires. It is easy to see that the above transformations preserve the location of the interface at $\tilde{x}_i=0$ and ensure that 
\begin{eqnarray}
    \tilde{t}_i = t_n, \quad i=1,\cdots, n-1
\end{eqnarray}
so that we have continuous coordinates and metric across any pair of wires.

After the conformal transformations, the non-vanishing components of the energy-momentum $T^i_\pm = \frac{\pi c}{12} T^2$ on the $i^{\rm th}$ wire transform to
\begin{equation}\label{Eq:Texact}
    \widetilde{T}_{\pm\pm}^i(\tilde{x}_i^\pm)= {(\mathbb{h}_i'(\tilde{x}_i^\pm))^2}\frac{\pi c T^2}{12}-\frac{c}{24\pi}{\rm Sch}(\mathbb{h}_i(\tilde{x}_i^\pm),\tilde{x}_i^\pm),
\end{equation}
for $i \neq n$, and for $i=n$ we have $\widetilde{T}^{n}_{\pm \pm} = T^n_{\pm\pm}$.

For concreteness, let the spatial $\tilde{x}$ coordinate of each wire extend from $-\infty$ to $0$. In order to formulate the Ward identity, it is useful to consider the $i^{\rm th}$ wire described by CFT$_i$ on the left side of the multi-interface and reflect the remaining $n-1$ wires described by the combined $\overline{\rm CFT}_1\otimes\cdots \overline{\rm CFT}_{i-1}\otimes \overline{\rm CFT}_{i+1}\otimes\cdots \overline{\rm CFT}_{n}$ theory living in the semi-infinite interval extending from $0$ to $\infty$ and with $\overline{\rm CFT}$ denoting the exchange of left and right movers in the CFT due to the reflection. In this continuous $\tilde{x}$ coordinates, the non-vanishing energy momentum tensor is then
\begin{align*}
  &\widetilde{T}_{++}(\tilde{t},\tilde{x}) = 
         \Theta(-\tilde{x})\widetilde{T}^i_{++}(\tilde{t},\tilde{x}) (\tilde{t},\tilde{x})+ 
         \Theta(\tilde{x})\sum_{j\neq i}\widetilde{T}^j_{--}(\tilde{t},\tilde{x}),\nonumber\\
   &\widetilde{T}_{--}(\tilde{t},\tilde{x}) = 
         \Theta(-\tilde{x})\widetilde{T}^i_{--}(\tilde{t},\tilde{x}) (\tilde{t},\tilde{x})+ 
         \Theta(\tilde{x})\sum_{j\neq i}\widetilde{T}^j_{--}(\tilde{t},\tilde{x})
\end{align*}

The Ward identities are
\begin{align}
    &\partial_{\tilde{t}}\tilde{T}^{\tilde{t}\tilde{t}}(\tilde{t},\tilde{x})+\partial_{\tilde{x}}\tilde{T}^{\tilde{x}\tilde{t}}(\tilde{t},\tilde{x}) =0,\label{Eq:WI1}\\
    &\partial_{\tilde{t}}\tilde{T}^{\tilde{t}\tilde{x}}(\tilde{t},\tilde{x})+\partial_{\tilde{x}}\tilde{T}^{\tilde{x}\tilde{x}}(\tilde{t},\tilde{x})={\delta(\tilde{x})}q(\tilde{t}).\label{Eq:WI2}
\end{align}
The first equation above follows as $T^{xt}\propto T_{++}- T_{--}$ and as evident from \eqref{Eq:Texact}, $\tilde{T}^k_{++}-\tilde{T}^k_{--}$ vanishes at $\tilde{x}=0$ for all $k =1,\cdots, n$. More generally, the right hand side of \eqref{Eq:WI1} should be proportional to
\begin{align}
    \sum_{i} \tilde{T}^i_{--} (\tilde{t}, \tilde{x} =0) - \sum_{i} \tilde{T}^i_{++} (\tilde{t}, \tilde{x} =0)
\end{align}
which vanishes if the conformal boundary condition is preserved as is indeed the case generally for the two-way junction (see \cite{Chakraborty:2025dmc}). The source appearing in \eqref{Eq:WI2},
\begin{align}
    &q(\tilde{t})=\sum_{j\neq i}\lb\tilde{T}^j_{++}+\tilde{T}^j_{--}\rb(\tilde{t},\tilde{x}=0) \nonumber\\&-\lb\tilde{T}^i_{++}+\tilde{T}^i_{--}\rb(\tilde{t},\tilde{x}=0),
\end{align}
is the expectation value of a generalized displacement operator which generates the displacement of the $i^{\rm th}$ wire away from the interface.

\subsection{Details of the perturbative expansion}
The details of the derivation of the Nambu-Goto-Monge-Amp\`{e}re equations \eqref{Eq:CoupledNGEOMs} from the $3n-2$ junction conditions via perturbative expansion in $\lambda$ are as follows.  At zeroth order, permutation symmetry implies trivial solutions $x_s = x_0$ and $x_{d_i} = \tau_{d_i} =\sigma_{d_i} =0$ of the metric continuity equations \eqref{Eq:hcont} and the condition for the extrinsic curvature discontinuity \eqref{Eq:Kdisc}.  At first order in $\lambda$, only the $\mathcal{O}(\epsilon)$ terms of $\tau_{d_i}$ and $\sigma_{d_i}$ appear in the $3n-3$ metric continuity equations \eqref{Eq:hcont}. Permutation symmetry implies $\tau_{d_i}$ and $\sigma_{d_i}$ should vanish at first order. At first order, the extrinsic curvature discontinuity \eqref{Eq:Kdisc} gives 
\begin{align}\label{Eq:Kdisc-order-1}
 \left(1 - M \sigma^2\right) \left( \lambda - n x^\prime_{s} \right) - n \sigma \ddot{x}_{s} =\mathcal{O}(\epsilon^2)
 \,, \\
  M \sigma \dot{x}_{s} + \left(1 - M \sigma^2\right)  \dot{x}^\prime_{s} = \mathcal{O}(\epsilon^2),, \\
 \left(1 - M \sigma^2\right)n \sigma x^{\prime \prime}_{s} - n x^\prime_{s} + \lambda =\mathcal{O}(\epsilon^2) \,.
\end{align}
which determines $x_s$ at the first order as $x_s = \lambda\sigma/n + \mathcal{O}(\epsilon^2)$ given the Dirichlet boundary condition \eqref{Eq:DBC}. This part is similar to the $2$-way junction case discussed in \cite{Banerjee:2024sqq}. 

Generally, $\mathcal{O}(\epsilon^{k})$ terms of $\tau_{d_i}$ and $\sigma_{d_i}$, and $\mathcal{O}(\epsilon^{k-1})$ term  of $x_{d_i}$ are determined at the $k$-th order in the perturbative expansion of the $3n-3$ metric continuity conditions \eqref{Eq:hcont} for $k\geq 2$. The $\mathcal{O}(\epsilon^{k})$ term of $x_{s}$ is determined at the $k$-th order of the expansion of the extrinsic curvature discontinuity condition \eqref{Eq:Kdisc} for $k\geq 1$. Also, the $\mathcal{O}(\epsilon^{k})$ terms of $x_{d_i}$ and $x_s$ vanish when $k$ is odd and even respectively. 

We generalize the procedure in \cite{Banerjee:2024sqq} to solve the $3n-3$ metric continuity conditions \eqref{Eq:hcont} at $k$-th order in the perturbative expansion for $k\geq2$. From the diagonal $\tau\tau$ and $\sigma\sigma$ components of the metric continuity conditions we can obtain $\tau_{d_i}$ and $\sigma_{d_i}$ explicitly in terms of $x_{d_i}$ up to terms which depend only on $\sigma$ or $\tau$. Substituting these forms of $\tau_{d_i}$ and $\sigma_{d_i}$ into the off-diagonal components of the metric continuity conditions yield $n-1$ equations that are schematically $\mathcal{E}_i =0$. Then $\partial_\sigma\partial_\tau(\mathcal{E}_i\sigma\sqrt{1-M\sigma^2})=0$ give the coupled Nambu-Goto-Monge-Amp\`{e}re equations \eqref{Eq:CoupledNGEOMs} involving \textit{only} the $n-1$ variables $x_{d_i}$. Given that $x_{d_i}$ satisfies these equations, we can determine  $\tau_{d_i}$ and $\sigma_{d_i}$ completely from $\partial_\tau\mathcal{E}_i =0$ and $\partial_\sigma(\mathcal{E}_i\sigma\sqrt{1-M\sigma^2})=0$. This procedure works at higher orders in the perturbative expansion also. 

For $k\geq 2$, the extrinsic curvature discontinuity takes the same linear form as \eqref{Eq:Kdisc-order-1} for the $\mathcal{O}(\epsilon^{k})$ term of $x_s$ only with source terms containing lower order terms of $x_s$ and other variables. These can be readily solved as in \cite{Banerjee:2024sqq}.

\subsection{A special case of Nambu-Goto-Monge-Amp\`{e}re equations}
The Nambu-Goto-Monge-Amp\`{e}re equations \eqref{Eq:CoupledNGEOMs} can be solved explicitly when the $n$-way junction can be reduced to the $2$-way junction. This reduction corresponds to setting $x_k = x_s$ for $k\neq i$ and $k\neq j$ (freezing the relative motions of $n-2$ of the $n$ hypersurfaces glued at the junction) which impose $n-2$ linear constraints on the $n-1$ variables $x_{d_i}$ so that only one linear combinations of the $x_{d_i}$ is independent. This linear combination satisfies the linearized Nambu-Goto equation \eqref{Eq:CNG} since the $n-2$ constraints imply that the quadratic Monge-Amp\`{e}re terms vanish identically. As for instance, setting $x_1 = x_3 = x_s$ in the case of the $4$-way junction impose that $x_{d_1} = 2 x_{d_2}$ and $x_{d_3} = x_{d_2}$, and the three coupled equations \eqref{Eq:CoupledNGEOMs} simply imply that  $x_{d_2}$ satisfies \eqref{Eq:CNG}. 

The normalizable solutions of the linearized Nambu-Goto equation \eqref{Eq:CNG} in the BTZ black hole background corresponding to the quasinormal modes of the string decay as $e^{-(2+m)\sqrt{M}\tau}$ for $m=0,1,2\cdots$   \cite{Banerjee:2024sqq}. Explicitly, the general solution is of the form
\begin{equation}
    x_{d_i} = \sum_{m= 0}^\infty b_m e^{-(2+m)\sqrt{M}\tau} P_{m}(\sigma),
\end{equation}
with
\begin{align}
    & P_{m}(\sigma) = \frac{3}{m^3 -m}\Big(\frac{m\,\sigma\cosh(m \, {\rm arctanh}(\sqrt{M}\sigma))}{M}\nonumber\\&\qquad\qquad -\frac{\sinh(m\,{\rm arctanh}(\sqrt{M}\sigma))}{M^{\frac{3}{2}}}\Big).
\end{align}
We can check that
\begin{align}
    & P_2(\sigma) = \frac{\sigma^3}{1-M\sigma^2},\,\, P_3(\sigma)= \frac{\sigma^3}{(1-M\sigma^2)^{3/2}},\,\,{\rm etc.}
\end{align}
As discussed in the main text, the normalizable solutions of the full coupled Nambu-Goto-Monge-Amp\`{e}re equations \eqref{Eq:CoupledNGEOMs} also decay as $e^{-(2+m)\sqrt{M}\tau}$ for $m=0,1,2\cdots$ when $\lambda\neq 0$.

\bibliographystyle{JHEP}
\bibliography{References}

\end{document}